\documentclass[12pt,a4paper]{article}
\usepackage{graphicx}
\usepackage{txfonts}
\usepackage{epstopdf}

\usepackage{longtable}
\usepackage{lscape}
\usepackage{natbib}

\usepackage{color}
\def\ALi{\hbox{$\thinspace A (\mathrm{Li})$}}

\def\aj{The Astronomical Journal}
\def\apjl{The Astrophysical Journal Letters}
\def\apj{The Astrophysical Journal}
\def\apjs{The Astrophysical Journal Supplement Series}
\def\aap{Astronomy and Astrophysics}
\def\aaps{ Astronomy and Astrophysics Supplement Series}
\def\mnras{Monthly Notices of the Royal Astronomical Society}
\def\pasp{ Publications of the Astronomical Society of Pacific}
\def\nat{Nature}
\def\apss{Astrophysics and Space Sciences}
\def\actaa{Acta Astronomica}
\def\pasj{Publications of the Astronomical Society of Japan}

\bibpunct{(}{)}{;}{a}{}{,} 

\begin{document}

  \title{Tracking Advanced Planetary Systems with HARPS-N (TAPAS).  I. A multiple planetary system around the red giant star TYC 1422-614-1.
  \thanks{Based on observations obtained with the Hobby-Eberly Telescope, which is a joint project of the University of Texas at Austin, the Pennsylvania State University, 
Stanford University, Ludwig-Maximilians-Universit\"at M\"unchen, and Georg-August-Universit\"at G\"ottingen.}
\thanks{Based on observations made with the Italian Telescopio Nazionale Galileo (TNG) operated on the island of La Palma by the Fundaci\'on Galileo Galilei of the INAF (Istituto Nazionale di Astrofisica) at the Spanish Observatorio del Roque de los Muchachos of the Instituto de Astrof\'{\i}sica de Canarias.}
}

\author{A. Niedzielski${}^1$, 
E. Villaver${}^2$,
A. Wolszczan${}^{3,4}$,
M. Adam\'ow${}^1$,
K. Kowalik${}^1$,\\
G. Maciejewski${}^1$,
G. Nowak${}^{5,6,1}$,
D. A. Garc\'{\i}a-Hern\'andez${}^{5,6}$,\\
B. Deka${}^1$,
M. Adamczyk${}^1$}

\footnotetext[1]{Toru\'n Centre for Astronomy, Faculty of Physics, Astronomy and Applied Informatics, Nicolaus Copernicus University in Toru\'n, Grudziadzka 5, 87-100 Toru\'n, Poland, e-mail: Andrzej.Niedzielski@umk.pl}
\footnotetext[2]{Departamento de F\'{\i}sica Te\'orica, Universidad Aut\'onoma de Madrid, Cantoblanco 28049 Madrid, Spain, e-mail: Eva.Villaver@uam.es}
\footnotetext[3]{Department of Astronomy and Astrophysics, Pennsylvania State University, 525 Davey Laboratory, University Park, PA 16802, USA.}
\footnotetext[4]{Center for Exoplanets and Habitable Worlds, Pennsylvania State University, 525 Davey Laboratory, University Park, PA 16802, USA.}
\footnotetext[5]{Instituto de Astrof\'isica de Canarias, E-38205 La Laguna, Tenerife, Spain.}
\footnotetext[6]{Departamento de Astrof\'isica, Universidad de La Laguna, E-38206 La Laguna, Tenerife, Spain.}

\date{Received ; accepted }

    \maketitle
    
  \abstract{

Context. 
{Stars that have evolved-off the Main Sequence  are crucial 
in
   expanding the frontiers of knowledge
   on exoplanets
  towards higher stellar
  masses, and to constrain star-planet interaction mechanisms.
These stars, however suffer from intrinsic activity that complicates
 the interpretation of precise radial velocity measurement and are often avoided in planet
  searches. We have, over the last 10 years, monitored about 1000 evolved
  stars for radial velocity variations in search for low-mass
companions under the Penn State - Toru\'n Centre for Astronomy
Planet Search  program with the Hobby-Eberly Telescope. Selected prospective candidates 
that required higher RV precision
meassurements have been followed with HARPS-N at the 3.6 m Telescopio Nazionale Galileo.
} 

Aims. 
{To detect planetary systems around evolved stars, to be able to build
  sound statistics on the frequency and intrinsic nature of these systems, and to deliver
  in-depth studies of selected planetary systems with evidences of
  star-planet interaction processes. } 

Methods. 
{We have obtained  for 
  TYC 1422-614-1 69 epochs of precise radial velocity
  measurements collected over 3651 days with the Hobby-Eberly
  Telescope, and 17 epochs of ultra precise HARPS-N data collected
  over 408 days. We have complemented these RV data with photometric
  time-series from the All Sky Automatic Survey archive.} 

Results.
{ We report the discovery of a multiple planetary system around the
  evolved K2 giant star TYC 1422-614-1. The system orbiting the 1.15
  M$_\odot$ star 
  is composed of a planet with mass
  m$sin i$=2.5 M$_J$ in a 0.69 AU orbit, and a planet/brown dwarf with
  m$sin i$=10 M$_J$ in a 1.37 AU orbit. The multiple planetary system
  orbiting TYC 1422-614-1 is the first finding of the
  \textit{TAPAS} project, a HARPS-N monitoring of
  evolved planetary systems identified with the Hobby-Eberly
  Telescope. }
{}

}

%

\section{Introduction}

Since Nicolaus Copernicus \citep{Copernicus1543} proposed for the
first time the model of the Solar System, it has taken nearly 450 yr to detect the first planet around stars
other than the Sun \citep{Wolszczan1992, MayorQueloz1995, 
  MarcyButler1996}. The progress has been much faster since, and
the first multiple planetary system orbiting a Main Sequence (MS) star 
was found only a few years later
by \cite{Butler_p_And_1999}. 

Nowadays, the focus of most planet searches is on MS stars with
planets in the stable Habitable Zones (HZ) and from a list of nearly
1\,800 exoplanets candidates\footnote{http://exoplanet.eu/}, 20 are
potentially habitable\footnote{http://phl.upr.edu/projects/habitable-exoplanets-catalog},
including the first Earth-like planet (Gliese 581 d -
\citealt{Udry2007}).  We have as well found extrasolar systems as
complex as our own (e.g. 55 Cnc  - \citealt{Fischer_55Cnc_2008}; HD 10180
- \citealt{Lovis2011}; Kepler 11 - \citealt{Lissauer2011}).   

Stars beyond the MS are frequently avoided in planet searches as they
are known to exhibit various types of variability: radial velocity  (RV) variations of
unknown origin were pointed out to be common in Red Giants (RGs) by \cite{Walker1989};
multiple pulsation modes are often present \citep{HatzesCochran1993, Wood1999,
  deRidder2009, Kallinger2010, Mosser2013}. Also 
rotation of starspots across the stellar disk can affect the spectral
line profiles of these stars \citep{Vogt1987, Walker1992, SaarDonahue1997}.

Several authors have collected precise data in order to understand the
origin of RV variations in RGs. Some of these RG stars with RV
variations ($\beta$ Gem - \citealt{HatzesCochran1993, 2006ApJ...652..661R, 2011A&A...527A.140R}; $\gamma$ Cep -
\citealt{Campbell1988, 2011AIPC.1331...88E, 2011A&A...527A.140R}) are now recognized as having indeed planetary-mass
companions \citep{Hatzes2003, Hatzes2006}.  HD 177830 \citep{Vogt2000},
and $\epsilon$ Ret
\citep{Butler2001} are interesting cases to
illustrate the problems with this type of systems given that both were reported
to host a planetary companion as ``evolved subgiants'' 
and only much later recognized to be  bona fide giants by
\cite{Mortier2013}. As a consequence,
 $\iota$ Dra b \citep{Frink2002} is usually considered the
first deliberate discovery of a planet around a RG star.  

Searches for planets around stars beyond the MS have, soon after the first
discovery, became recognized as
important in building a complete picture of planet formation and
evolution for several reasons.  First, they allow to extend the
reach of  the most versatile RV technique--not applicable on
the MS due to the high effective temperature of the stars and their
fast rotation rates--to objects with masses
significantly larger than solar (e.g $o$ UMa, a 3 M$_{\odot}$ giant
with a planet - \citealt{Sato2012}).  
Second, the planetary systems around evolved stars are much older
than those around MS stars and therefore suitable for long term
dynamical stability considerations \citep{debes02,verasmustill13,Mustill2014}. Planetary
systems around giants are also subject to changes induced by stellar
evolution \citep{VillaverLivio2007, VillaverLivio2009, Villaveretal14}, and therefore
suitable to studies of star - planet interactions
(e.g. \citealt{Adamow2012}), and last but not least, evolved planetary systems carry information on
 the initial population of planetary systems to be found around White Dwarfs
 \citep{Farihi2010}.  

It is no surprise then that several projects devoted to RV planet searches orbiting
RGs were launched: McDonald Observatory Planet Search
\citep{CochranHatzes1993, HatzesCochran1993},  Okayama Planet Search
\citep{Sato2003}, Tautenberg Planet Search \citep{Hatzes2005},  
Lick K-giant Survey \citep{Frink2002},  
ESO Ferros planet search \citep{SetiawanHatzes2003, SetiawanPasquini2003}, 
Retired A Stars and Their Companions \citep{Johnson2007}, 
Coralie \& HARPS search \citep{LovisMayor2007}, 
Boyunsen Planet Search \citep{Lee2011},  and one of the largest, the PennState - Toru\'n Centre
for Astronomy Planet Search (PTPS, \citealt{Niedzielski2007, Niedzielski2008, 
  NiedzielskiWolszczan2008}). 

Within PTPS 
we monitored over 1000 stars for RV
variations with the Hobby-Eberly Telescope  and its High Resolution
Spectrograph  since 2004.  As part of this effort about 300 planetary/Brown Dwarf (BD) candidates were
identified that call for more intense precise RV follow-up. TAPAS (Tracking
Advanced Planetary Systems with HARPS-N) is the result of intensifying
the monitoring of a selected number of PTPS identified targets, i.e. 
those with potentially  multiple and/or low-mass companions, with
expected p-mode oscillations of few m s$^{-1}$, as well as
systems with  evidence of recent or future star-planet interactions
(Li-rich giants, low orbit companions etc.).

TYC 1422-614-1 is a Red Clump Giant star selected as a TAPAS target 
since the HET observations showed that it was a relatively inactive giant with
expected p-mode oscillations of only 4 m s$^{-1}$ showing a complex 
RV variation pattern that called for more epochs of precise
data. After several years of PTPS observations our 
continuously updated model for the rare multi-planetary system around
this giant was still not consistent and since the very important and
rare RV minimum was expected during the period of HET not being
operational due to upgrade we chose to add this star to the TAPAS
target list.  

In this paper we report the discovery of a multiple planetary system
orbiting the giant star TYC 1422-614-1, the first planetary system result of TAPAS, our intensive
monitoring with HARPS-N of some of the PTPS selected targets. 
The paper is organized as follows: in
Section \ref{observations} we present the observations obtained for
this target and we outline the reduction and
measurements procedures; Section \ref{results-g} shows the results of
the Keplerian and Newtonian data modeling; in Section
\ref{activity} we extensively discuss the stellar activity influence
on the RV variation measurements; and in Section
\ref{discussion} and \ref{conclusions}, we discuss the results of our
analysis and present the conclusions. 

\section{Observations and data reduction. \label{observations}}

TYC 1422-614-1 (2MASS J10170667+1933304) is a V=10.21 and B-V=0.95 mag
\citep{Hog2000} star in the constellation of Leo. Since a trigonometric parallax is not
available for this star, its mass, age, and
radius  were estimated on the basis of spectroscopically determined
atmospheric parameters by \cite{Zielinski2012}. To provide a better constraint 
to the stellar parameters we have constructed a probability distribution function using  the algorithm of 
\cite{daSilva2006}, a modified version of the Bayesian estimation
method idealized by \cite{JorgensenLindegren2005} and
\cite{Nordstrom2004},  
and stellar isochrones from PARSEC (PAdova and TRieste Stellar Evolution Code, \citealt{Bressan2012}). 
Even though our results agree very well with those of
\cite{Zielinski2012}  without a trigonometric parallax they are
rather uncertain and model dependent. The amplitude of p-mode
  oscillations V$_{osc}$ was estimated from the scaling relation by \cite{KjeldsenBedding1995}.
The v$_{rot}^{CCF}$ $\sin i_{\star}$ was obtained by \cite{Nowak2012} from Cross-Correlation Function (CCF) analysis of
about 200 spectral lines following the prescription of
\cite{Carlberg2011}.    \cite{Adamow2014}, based on
abundance calculations  via SME \citep{SME1996} modeling of 27 spectral lines for  6
elements,  found no chemical
anomalies for TYC 1422-614-1 with respect to a full
sample of RG stars. A summary of all the available data for TYC 1422-614-1 is given in Table \ref{Parameters}.

\begin{table}
\centering
\caption{Summary of the available data on TYC 1422-614-1.}
\begin{tabular}{lll}
\hline
Parameter & value & reference\\
\hline\hline
V  [mag]& 10.21& \cite{1997ESASP1200.....P} \\
B-V [mag] & 0.95$\pm$0.085 & \cite{1997ESASP1200.....P} \\
(B-V)$_0$ [mag] & 0.997 & \cite{Zielinski2012} \\
M$_V$ [mag] & 0.81 & \cite{Zielinski2012} \\
 T$_{eff}$ [K] & 4806$\pm$45 & \cite{Zielinski2012} \\
 logg & 2.85$\pm$0.18& \cite{Zielinski2012} \\
$[Fe/H]$ & -0.20$\pm$0.08 & \cite{Zielinski2012}\\
RV [kms$^{-1}$] & 37.368$\pm$0.027& \cite{Zielinski2012} \\
v$_{rot}^{CCF}$ $\sin i_{\star}$ [kms$^{-1}$] & 1.4$\pm$0.7 & \cite{Nowak2012}\\
$\ALi $& $<1.1$ &  \cite{Adamow2014} \\
\hline
\hline
M/M$_{\odot}$ & 1.15$\pm$0.18 & this work\\
log(L/L$_{\odot}$) & 1.35$\pm$0.16 & this work\\
R/R$_{\odot}$ & 6.85$\pm$1.38& this work\\
log age [yr]& 9.77$\pm$0.22& this work\\
d [pc] & 759 $\pm$ 181 & calculated from M$_{V}$\\
V$_{osc}$ [ms$^{-1}$] & 4.555$^{+3.718}_{-1.993}$ & this work\\
P$_{osc}$ [d] & 0.141$^{+0.102}_{-0.064}$ & this work\\
\hline
\end{tabular}
\label{Parameters}
\end{table}


The spectroscopic observations presented in this paper were made  with
the 9.2 meter effective aperture  (11.1 x 9.8m) Hobby-Eberly Telescope
(HET, \citealt{Ramsey1998}) and its  High Resolution Spectrograph (HRS,
\citealt{Tull1998}) in the queue scheduled mode \citep{Shetrone2007},
and with the 3.58 meter Telescopio Nazionale Galileo (TNG) and its High
Accuracy Radial velocity Planet Searcher in the North hemisphere (HARPS-N,
\citealt{Cosentino2012}). Time-series of photometric data were obtained
from the All Sky Automated Survey (ASAS, \citealt{Pojmanski2002}). 

\subsection{Hobby-Eberly Telescope data}

The HET \& HRS spectra used in this paper were gathered with the HRS fed
with a 2 arcsec fiber, working in the R=60\,000 mode with a gas cell
($I_2$) inserted into the optical path. The spectra consisted of 46
echelle orders recorded on the ``blue'' CCD chip (407.6-592 nm) and 24
orders on the ``red'' one (602-783.8 nm).  
Details of our survey, the observing procedure, and data analysis
 have been described in detail elsewhere \citep{Niedzielski2007,
    Nowak2013}. The configuration and observing
 procedure employed in our program were, in practice, identical to those
 described by \cite{Cochran2004}.  
The basic data reduction was performed using standard IRAF\footnote{IRAF is distributed by the National Optical Astronomy Observatories, which are operated by the Association of Universities for Research in Astronomy, Inc., under cooperative agreement with the National Science Foundation.} tasks and scripts developed for PTPS.  
With the  precision reached we use the \cite{Stumpff1980} algorithm to
refer the measured RVs to the Solar System barycenter.

\begin{center}
\begin{longtable}{lrrrr}
\hline
\caption[]{HET \& HRS RV and BS measurements of  TYC 1422-614-1.} \label{HETdata} \\
\hline
\hline

MJD& RV [ms$^{-1}$] & $\sigma_{RV}$ [ms$^{-1}$] & BS [ms$^{-1}$] & $\sigma_{BS}$ [ms$^{-1}$]\\
\hline

\endhead

\endfoot

\hline \hline
\endlastfoot

 53034.254184  &    201.61   &    10.91  &      7.15  &     17.79  \\  
 53341.420903  &   -277.82   &     8.86  &    -40.75  &     10.281  \\   
 54060.455116  &     58.74   &     7.21  &     10.89  &     18.62  \\    
 54198.278600  &    174.18   &     7.12 &     30.91  &     19.80  \\    
 54216.230660  &    101.99   &     6.78  &     -0.41  &     19.81  \\  
 54254.129931  &   -109.79   &     8.00  &      7.19  &     22.56  \\    
 54516.208681  &   -109.91   &     7.30  &    -19.29  &     18.88  \\    
 54575.245828  &    142.40   &     8.78  &      5.31  &     10.46  \\    
 54619.123189  &    212.62   &     9.21  &     70.09  &     26.96  \\    
 54775.490203  &     80.91   &     8.45  &     -4.51  &     26.53  \\    
 54792.437269  &     78.37   &     7.70  &     30.43  &     21.23  \\    
 54811.393223  &     14.82   &     9.74  &      6.40  &     29.30  \\    
 54837.316163  &   -115.27   &    11.67  &     -0.31  &     40.95  \\    
 54837.329063  &   -107.42   &    12.38  &     75.23  &     39.54  \\    
 54845.509659  &   -140.33   &     9.77  &     -7.47  &     27.24  \\    
 54874.429225  &   -269.11   &     7.79  &     -4.04  &     19.48  \\    
 55184.362749  &    192.26   &     8.63  &     30.85  &     32.19  \\    
 55184.374468  &    197.96   &     8.79  &      0.83  &     26.31  \\    
 55209.303594  &    207.83   &     7.12  &     17.39  &     23.89  \\    
 55209.313466  &    209.53   &     7.64  &      4.82  &     26.16  \\    
 55218.292894  &    220.24   &     9.00  &    -28.27  &     27.17  \\    
 55222.270272  &    187.358   &     6.90  &      0.13  &     19.40  \\    
 55268.350168  &     80.33   &     7.46  &     49.10  &     15.92  \\    
 55502.500185  &   -371.49   &     7.78  &    -17.98  &     22.42  \\    
 55518.451788  &   -361.94   &     7.62  &     -4.79  &     22.18  \\    
 55527.439971  &   -384.03   &     6.66  &     16.32  &     19.70  \\    
 55566.339595  &   -213.77   &     8.89 &      7.95  &     28.54  \\    
 55603.224138  &    -97.50   &     6.86  &     -5.88 &     17.14  \\    
 55634.139728  &    -70.81   &     6.40  &     67.60  &     14.07  \\    
 55688.206001  &    -69.46   &     6.75  &     38.01  &     17.86  \\    
 55883.448738  &    -49.99   &     8.23  &     18.67  &     23.76  \\    
 55930.323171  &    -68.38   &     6.94  &     28.96  &     20.55  \\    
 55956.462627  &   -114.70   &     6.25  &     30.60  &     17.49  \\    
 55971.420764  &   -104.14   &     6.25  &      8.96  &     14.53  \\    
 55987.377512  &   -129.70   &     6.56  &     66.70  &     18.15  \\    
 55997.344537  &   -149.67   &     7.57  &    -28.59  &     21.90  \\    
 56009.318727  &   -189.08   &     6.06  &     64.38  &     16.28  \\    
 56016.287488  &   -217.76   &     6.38  &     17.15  &     15.94  \\    
 56018.286516  &   -225.42   &     7.01  &    -10.45  &     16.14  \\    
 56060.165961  &   -378.08   &     8.53  &     36.02  &     25.98  \\    
 56063.178310  &   -379.03   &     6.59  &     40.91  &     18.01  \\    
 56071.135023  &   -380.66   &     8.36  &     42.97  &     25.69  \\    
 56089.113461  &   -381.59   &     7.29  &     10.16  &     18.20  \\    
 56232.501065  &    -47.09   &     6.41  &     -6.83  &     13.74  \\    
 56243.470914  &    -19.80   &     6.80  &     -2.60  &     15.24  \\    
 56255.443970  &    -24.33   &     6.71  &     22.27  &     16.33  \\    
 56259.432107  &    -23.30   &     5.75  &    -12.58  &     13.85  \\    
 56262.420046  &    -10.08   &     6.55  &    -10.74  &     16.13  \\    
 56280.380990  &     11.011  &     7.99  &     46.65  &     25.03  \\    
 56313.288785  &     98.37   &     5.94  &     10.91  &     15.92  \\    
 56327.437049  &    137.30   &     6.89  &      8.93  &     16.34  \\    
 56345.183628  &    213.16   &     7.57  &     53.46  &     19.54  \\    
 56354.162980  &    230.34   &     6.54  &     67.84  &     17.30  \\    
 56359.138568  &    207.75   &    10.23  &    -49.80  &     38.27  \\    
 56360.131946  &    228.97   &    11.18  &    -70.25  &     42.11  \\    
 56363.341221  &    231.11   &     6.84  &    -17.48  &     20.15  \\    
 56372.132506  &    226.68   &     6.96  &     42.44  &     17.91  \\    
 56378.319306  &    231.26   &     7.67  &     16.70  &     27.31  \\    
 56379.088264  &    228.93   &     6.52  &      3.81  &     19.32  \\    
 56379.289300  &    209.79   &    11.91  &    -20.05  &     40.86  \\    
 56381.280081  &    237.11   &     6.44  &      6.18  &     16.45  \\    
 56395.262373  &    220.22   &     6.89  &     16.10  &     16.08  \\    
 56408.226892  &    174.33   &     8.13  &     37.06  &     23.73  \\    
 56410.220920  &    184.51   &     6.85  &     49.91  &     15.98  \\    
 56416.205469  &    160.85   &     7.12  &     30.07  &     21.18  \\    
 56425.181013  &    142.44  &     6.73  &     22.24  &     16.01  \\    
 56431.159358  &     99.29   &     6.71  &     36.12  &     16.48  \\    
 56444.120550  &     18.04   &     6.77  &     27.00  &     17.04  \\    
 56445.128015  &     23.41   &     7.16  &      1.61  &     18.03  \\

\end{longtable}
\end{center}

Since HRS is a general purpose spectrograph, neither temperature nor
pressure controlled, the precise  RV measurements with this instrument
are best accomplished with the I2 cell technique. We use a combined
gas-cell  \citep{MarcyButler1992, Butler1996}, and  cross-correlation
\citep{Queloz1995, Pepe2002} method for this purpose.  
The implementation of this technique to our data is described in 
 \cite{Nowak2012} and \cite{ Nowak2013}. 
The precision of RV and line bisectors velocity span (BS), as well as the
long-term stability  of our measurements, has been verified by the
analysis of data obtained from the monitoring of stars that do not
exhibit detectable RV variations and stars with well described RV or
BS variations. The results for the K0 giant BD+70 1068
presented in \cite{Niedzielski2009b} show a $\sigma$=12 m s$^{-1}$, to
which contributed an intrinsic RV uncertainty  of 7 m s$^{-1}$  
and, the approximately 10 m s$^{-1}$ amplitude of solar-type 
oscillations \citep{KjeldsenBedding1995}. In \cite{Nowak2013} we performed
a more detailed analysis of our RV precision and stability using our own
RVs to fit orbital solutions for  HD~209458 and HD~88133, while our BS precision was demonstrated for  HD~166435. In general, we find
that our observing procedure with HET \& HRS results in RV
precisions of the order of 5-8 m s$^{-1}$ depending on the 
sigma-to-noise ratio (SNR) of the spectra, and the effective
temperature of the star, for BS measurements the precision is 2-3 times lower. 

This instrumental setup, observing, and reduction techniques and data
modeling tools have allowed us to report the discovery of 19
planetary mass companions, mostly to evolved stars
(\cite{Niedzielski2007, Niedzielski2009a, Niedzielski2009b,
  Gettel2012b, Gettel2012a,  Adamow2012, Nowak2012, Nowak2013}; Niedzielski et al. - in prep.)

\subsection{HARPS-N data}

HARPS-N, a near-twin of the HARPS instrument mounted
at the ESO 3.6-m telescope in La Silla \citep{Mayor2003}  is an echelle spectrograph covering the visible
wavelength range between 383 and 693 nm with a resolving power of $\mathrm{R}\sim115\;000$. 
The spectra are re-imaged on a 4kx4k CCD, where echelle spectra of 69 orders are formed for each fiber.
 
The instrument  is located in a thermally-controlled
environment, within a vacuum-controlled enclosure to ensure the
required stability, and is fed by two octagonal fibres at the Nasmyth B
focus of the TNG (with an aperture on
the sky of 1 arcsec). 
Both fibers are equipped with an image scrambler to provide a uniform
spectrograph pupil illumination, independent of pointing decentering. 
This instrument is expected to deliver measurements of radial
velocities with the highest accuracy currently available, with 1 m
s$^{-1}$ achievable from SNR=100 spectra. 

RV measurements, and their uncertainties, as well as BS were obtained
with the standard user pipeline which is based on the weighted
CCF method \citep{1955AcOpt...2....9F, 1967ApJ...148..465G, 1979VA.....23..279B, Queloz1995, Baranne1996, 
  Pepe2002}. For maximum precision of RV we used the simultaneous Th-Ar calibration 
mode of the spectrograph. TYC 1422-614-1 RVs were obtained with K5 
cross-correlation mask. 

The instrument has already successfully used in the detection of
exoplanets by \cite{Covino2013, Desidera2013, Hebrard2013} and
\cite{Pepe2013}.

\begin{table}
\centering
\caption{TNG \& HARPS-N RV and BS measurements of  TYC 1422-614-1.}
\begin{tabular}{lrrr}
\hline
MJD& RV [kms$^{-1}$] & $\sigma_{RV}$ [kms$^{-1}$] & BS [kms$^{-1}$] \\
\hline\hline

  56277.214162    &    37.76753   &   0.00120    &   0.07582 \\
  56277.286630    &    37.76889   &   0.00160    &   0.07580 \\
  56294.098292    &    37.77515   &   0.00094    &   0.08609 \\
  56294.275817      &     37.79236     &    0.00108      & 0.07573   \\
  56321.071935    &    37.87748   &   0.00153    &   0.08774 \\
  56321.229399    &    37.88666   &   0.00128    &   0.08015  \\
  56373.919742  &    37.96977 &   0.00122  &     0.08078 \\
  56374.050295  &    37.98511 &    0.00168  &      0.08346 \\
  56410.873491  &     37.93558 &     0.00143  &      0.10052 \\
  56411.019144  &     37.91875 &    0.00146  &      0.10148 \\
  56430.900812  &     37.81872 &    0.00179  &     0.08952 \\
  56430.977531  &     37.84129 &    0.00146  &     0.08561 \\
  56647.145306  &     37.41658 &    0.00257  &     0.08056 \\
  56647.275526  &     37.43739 &    0.00175  &     0.07915 \\
  56685.050558  &     37.43522 &    0.00169  &     0.07766 \\
  56685.200954  &     37.45209 &    0.00197  &     0.06340 \\
  56685.268427  &     37.43200 &    0.00280  &     0.06131 \\

\hline
\end{tabular}
\label{HARPSdata}
\end{table}


\section{Modeling of RV  data}\label{results-g}

TYC 1422-614-1 was observed over 3651 days between MJD= 53034 and
56685. In total, we collected 86 epochs of precise RV and BS, 69 with
HET and 17 with HARPS-N. 

The data obtained with HET \& HRS covers 69 epochs of observations over 3410 days
between MJD 53034 and MJD 56445 with typical SNR of 150-250 at 569
nm. The best quality template with SNR of 316 was used for the RV
measurements. The resulting RVs show an peak-to-peak amplitude of 621.1 ms$^{-1}$ and
uncertainties of 7.7 ms$^{-1}$ on average. The BS peak-to-peak amplitude is 145
ms$^{-1}$, with a mean value of 14.1 ms$^{-1}$ and  a precision of 21.3
ms$^{-1}$ on average (see Table \ref{HETdata}). The observed RV variations
are $\sim$4 times higher than the BS variations and  are not correlated (r=0.005)
hence they are most likely due to Doppler shifts. 

We also collected 17 epochs of RV with HARPS-N over 408 days between
MJD  56277 and MJD  56685 with an average uncertainty of 1.6 ms$^{-1}$ 
(see Table \ref{HARPSdata}).  HARPS-N  RV measurements show an peak-to-peak amplitude of 568.5
ms$^{-1}$. The BS  shows a mean value of 82.5 ms$^{-1}$ and peak-to-peak amplitude of
40.1 ms$^{-1}$, i.e 15 times lower than RV. The RV and BS show a
correlation of r=0.63 which is marginally significant as the critical
value is r$_{15,0.01}$=0.61. A closer look at the BS data shows that
this is likely a random effect due to small statistics and we assume
that at least in the first approximation the RV signal is indeed due
to Doppler shifts (we will come back to this issue in Section
\ref{activity}). 

 \begin{figure}[p]
   \centering
 \includegraphics[width=0.75\textwidth]{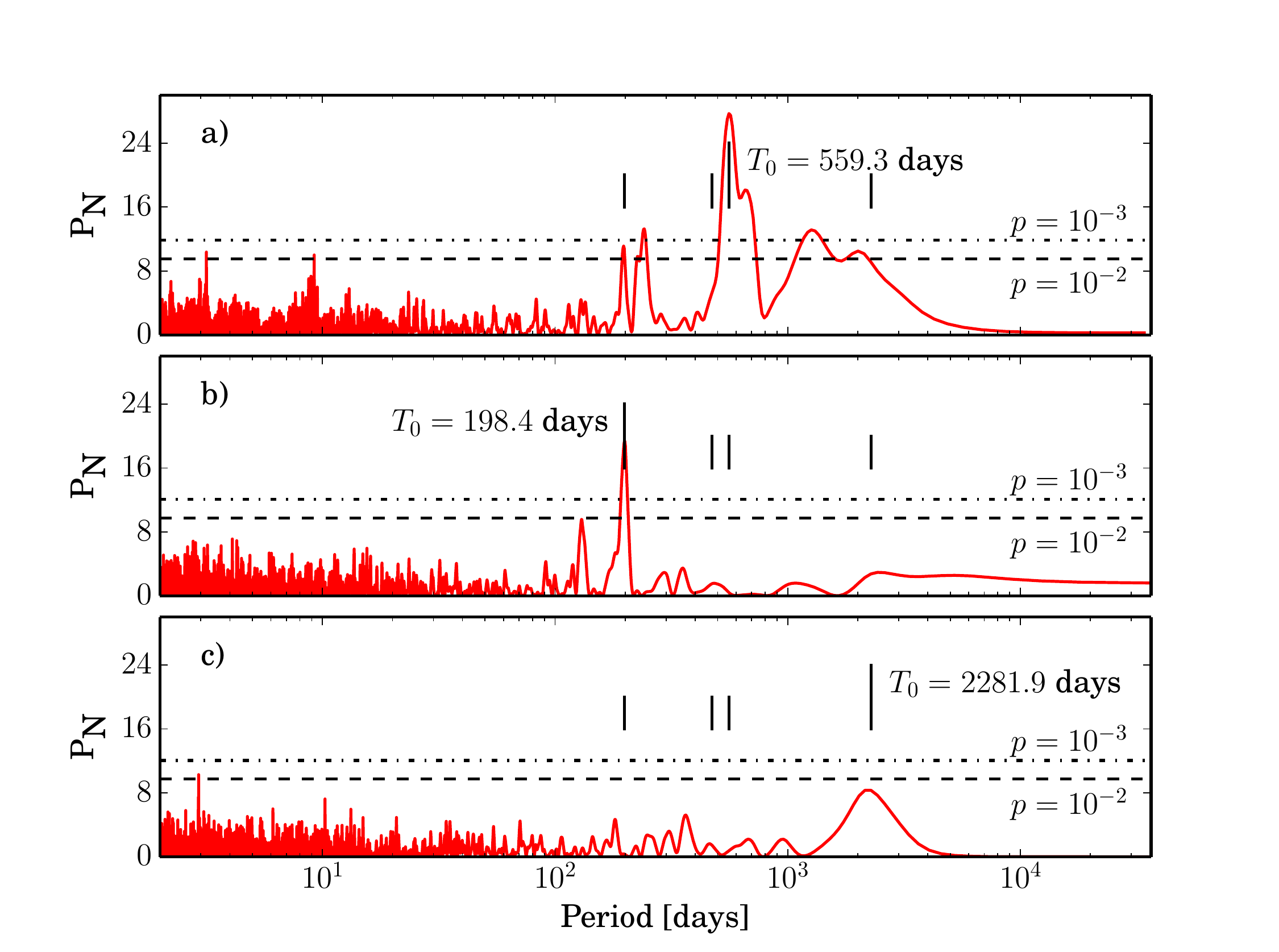}
      \caption{From top to bottom Lomb-Scarge periodograms for the
        (a) original HET RV data of TYC 1422-614-1,  (b) RV residua (HET and TNG) with planet c
        removed,  and (c) RV residua (HET and TNG) after the best Keplerian two-planetary fit. See also Figure \ref{LS-ph}.} 
         \label{LS}
         \end{figure}

\subsection{Keplerian analysis \label{SecKepler}}

We modeled Keplerian orbits to the observed RV variations using the
hybrid approach proposed by
\cite{Gozdziewski2003,GozdziewskiMigaszewski2006}, and \cite{Gozdziewski2007}.
The global search for  orbital parameters was done with a genetic
algorithm (GA). After the range of plausible orbital parameters was
narrowed,  a non-linear least squares fit to the data quickly converge to
the final 
solution.  In particular, with a set of periodic signals in RV time series
identified in its Lomb-Scargle (LS) periodogram
\citep{Lomb1976,Scargle1982}, we searched for possible orbital
solutions over a wide range of parameters with PIKAIA
\citep{Charbonneau1995}. We present the LS periodogram of TYC 1422-614-1 RV in Figure \ref{LS} (top panel). 
The GA semi-global search identified a narrow parameter 
range in the search space, which was then explored using the MPFIT 
algorithm  \citep{Markwardt2009} to locate the best-fit Keplerian
solution delivered by RVLIN \citep{WrightHoward2009}
modified after \cite{2011ApJS..197...26J} to allow the stellar jitter to be fitted as a free parameter. 

The scrambling of residuals method, or "bootstrapping"  \citep{1993ApJ...413..349M, 1997A&A...320..831K, Marcy2005,
  Wright2007} was used to assess the uncertainties of the best-fit orbital
parameters.  
The width of the resulting distribution of 10$^5$ trials between the
15.87th and 84.13th percentile is adopted as a parameter uncertainty.

\begin{figure}[p]
   \centering
   \includegraphics[width=0.75\textwidth]{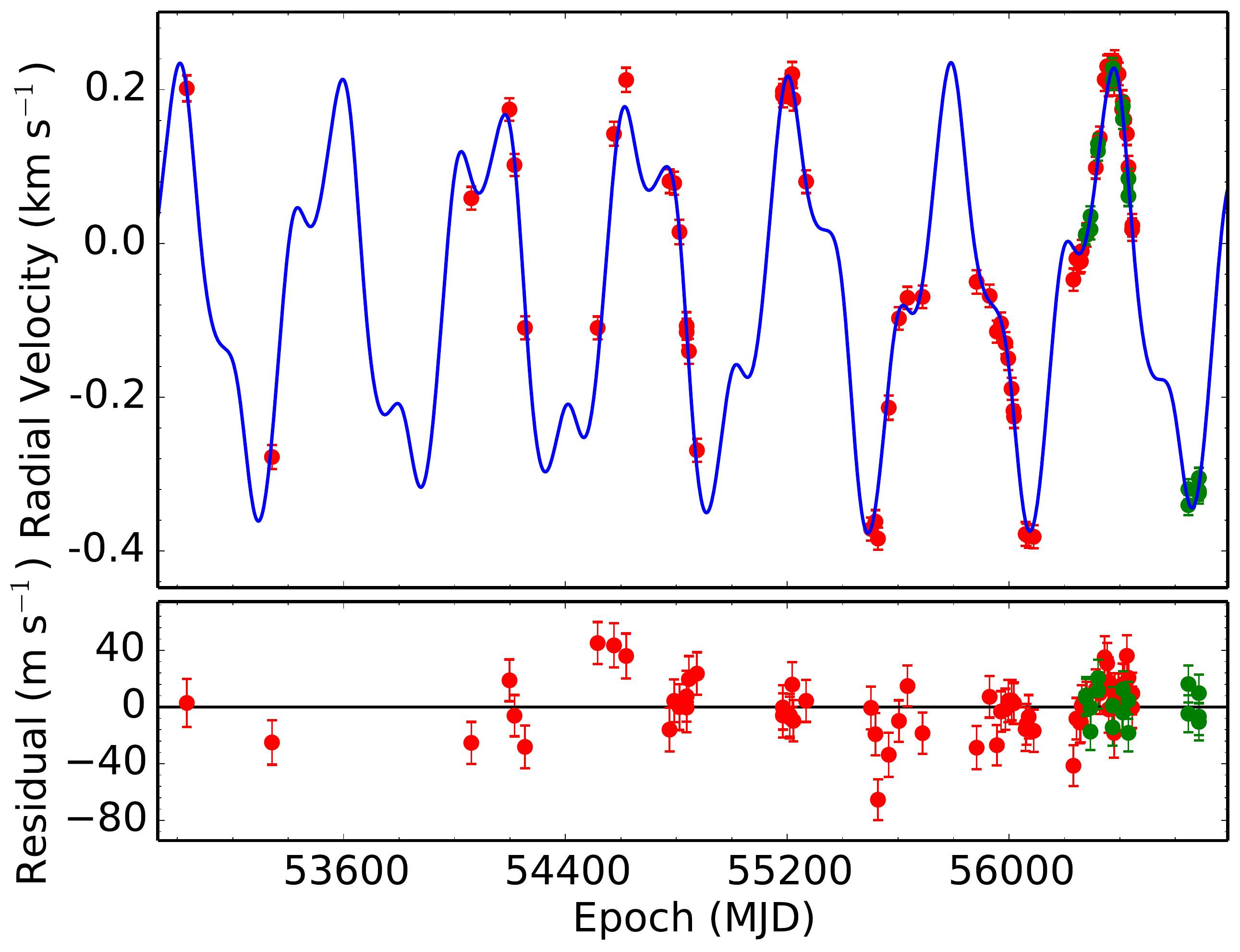}
      \caption{Keplerian best fit of the two-planetary model to the
        observed RV of  TYC 1422-614-1.  HET \& HRS data
        are presented in red and TNG \& HARPS-N data in green. The
        estimated jitter due to p-mode oscillations has been added to the
        uncertainties. } 
         \label{BestFit}
         \end{figure}

\begin{figure}[p]
   \centering
   \includegraphics[width=0.75\textwidth]{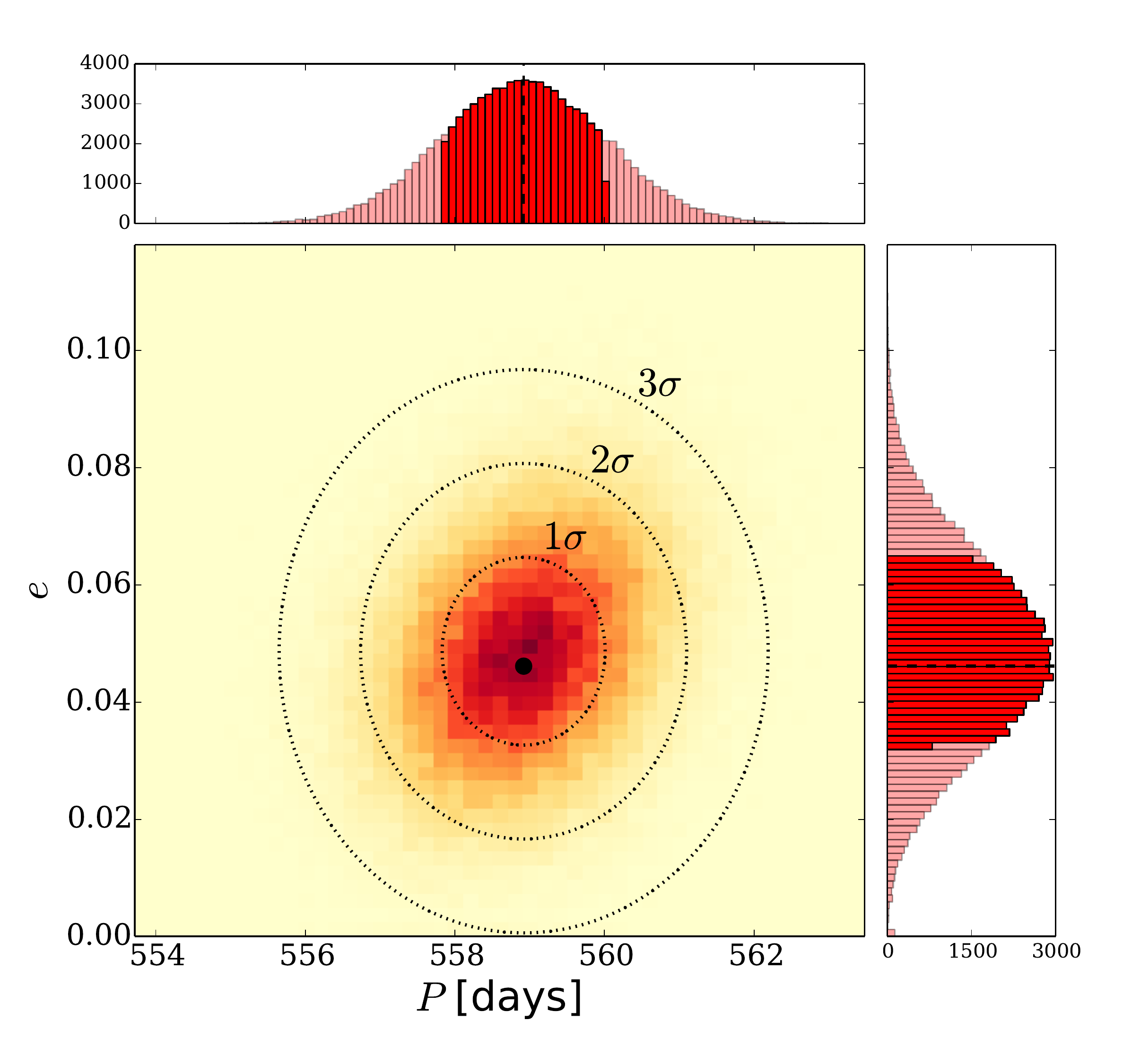}
      \caption{Illustration  of the bootstrapping analysis (10$^5$
        realizations) of the Keplerian parameters of TYC 1422-614-1
        c uncertainties, a  2D histogram of orbital period and eccentricity. Both 1D
        histograms are plotted with the 15.87 and 83.13 percentile
        (i.e. 1 $\sigma$) ranges shown. } 
         \label{Hist2D}
         \end{figure}

Results of the Keplerian analysis are presented in Figure \ref{BestFit} and Table \ref{KeplerianFit}. 
As an example,  the bootstrapping analysis  of orbital
parameters for component c is presented in Figure  \ref{Hist2D} where
we present a 2D period-eccentricity distribution.  
 
The false alert probability FAP$<$0.0005 of the final orbital solution was
estimated by repeating the whole hybrid Keplerian analysis described
above to 20\,000  sets of scrambled data.

 \begin{table}
\centering
\caption{Keplerian orbital parameters of TYC 1422-614-1 b and c.}
\begin{tabular}{lll}
\hline
Parameter & TYC 1422-614-1 b & TYC 1422-614-1 c \\
\hline
\hline
$P$ (days)                                        & $198.40^{+0.42}_{-0.42}$ & $559.3^{+1.2}_{-1.2}$ \\
$T_0$ (MJD)                                       & $53236^{+25}_{-22}$      & $53190^{+30}_{-30}$ \\
$K$ ($\textrm{m s}^{-1}$)                         & $82.0^{+7.0}_{-5.1}$     & $233.0^{+4.5}_{-4.0}$ \\
$e$                                               & $0.06^{+0.06}_{-0.02}$   & $0.048^{+0.020}_{-0.014}$ \\
$\omega$ (deg)                                    & $50.0^{+50}_{-43}$       & $130^{+20}_{-20}$ \\
$m_2\sin i$ ($\textrm{M}_{\textrm{J}}$)           & $2.5  \pm 0.4$           & $10  \pm 1$ \\
$a$ (AU)                                          & $0.69 \pm 0.03$          & $1.37 \pm 0.06$ \\
$V_0$ ($\textrm{m s}^{-1}$)                       & \multicolumn{2}{c}{ $-68.2^{+2.0}_{-2.2}$ } \\
offset ($\textrm{m s}^{-1}$)                      & \multicolumn{2}{c}{ $37758^{+6.0}_{-6.0}$ } \\
$\sigma_{\textrm{jitter}}$ ($\textrm{m s}^{-1}$)  & \multicolumn{2}{c}{ $12.9^{+1.4}_{-1.2}$ } \\
$\sqrt{\chi_\nu^2}$                               & \multicolumn{2}{c}{ $1.64$ } \\
$\sigma_{\textrm{RV}}$ ($\textrm{m s}^{-1}$)      & \multicolumn{2}{c}{ $18.94$ } \\
$N_{\textrm{obs}}$                                & \multicolumn{2}{c}{ $86$ }\\
\hline
\end{tabular}
\label{KeplerianFit}
\end{table}

The resulting ``jitter'' is larger than our estimate of unresolved p-mode oscillations
 of 4 m s$^{-1}$
and points to an additional source of RV
 scatter of unknown nature, to an additional component, or to stellar activity.

The keplerian  2-planetary best fit leaves behind a marginally
significant period of 2.9 days and a weaker one of $\sim$2280 days
(Figure \ref{LS}). The short period is probably an artifact, as the
fundamental radial pulsational period which we estimate according to
the formalism of \cite{Cox1972} is expected to be much shorter, 0.52 days.
The longer period is very uncertain and comparable with the timespan of
our observations, if confirmed with future observations may turn out
to be  due to another, yet unresolved  companion or long-term stellar
variability.

\subsection{Newtonian analysis \label{SecNewton}}

The ratio between orbital period of both planets is $P_{\rm{c}}/P_{\rm{b}}=0.355$, close to a 7:20 commensurability, what suggests that the system could be in a mean motion resonance. A Newtonian model of the system was obtained with the Systemic 2.16 software \citep{Meschiari2009}. The Keplerian orbital solution provided the initial parameters for the Bulirsch--Stoer integrator, running with a precision parameter of $10^{-16}$. The planetary system was assumed to be co-planar. The differential evolution algorithm with 5000 steps and Lavenberg-Marquardt minimization were used to find the best-fitting model. The bootstrap method with $10^5$ trials was used to estimate parameter uncertainties, calculated as median absolute deviations. The parameters of the best-fitting dynamical model are given in Table~\ref{Tab.DynResults}. In most cases, they agree with the Keplerian model well within 1$\sigma$. The only exception is $P_{\rm{c}}$, for which the dynamical solution gives the value greater by $\sim$$10$ days. The goodness of the fit was found to be similar to that of the Keplerian approach.  

\begin{figure}[p]
   \centering
 \includegraphics[width=0.75\textwidth]{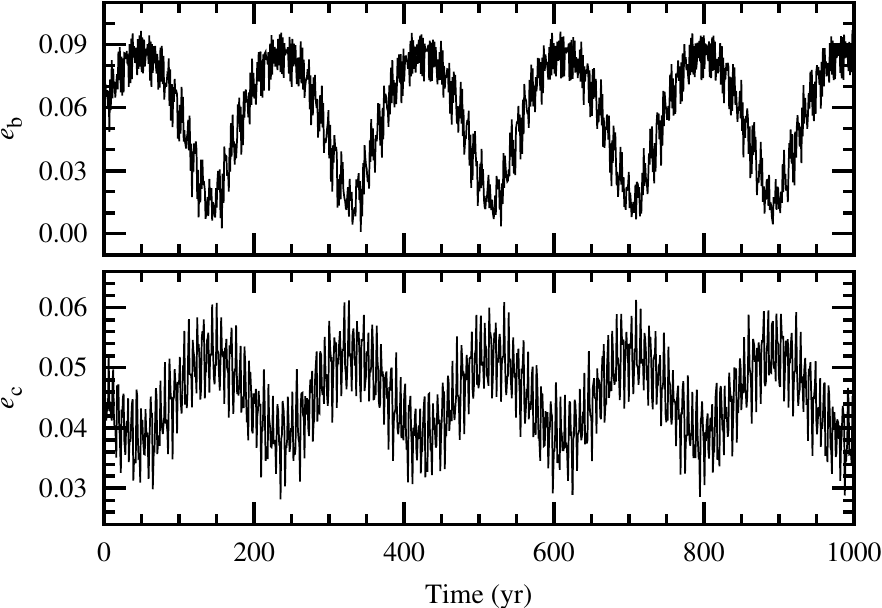}
      \caption{Evolution of planetary eccentricities in the first thousand years, calculated for the orbital solution given in Table~\ref{Tab.DynResults}} 
      \label{Fig.DynVariations}
         \end{figure}

The planetary system was found to be stable in a timescale of $10^6$ yr. The SWIFT's Regularized Mixed Variable Symplectic (RMVS) integrator was used to trace the behavior of orbital parameters. The simulation shows that $a_{\rm{b}}$ may vary between 0.685 and 0.693 AU  and $e_{\rm{b}}$ may oscillate between 0.0 and 0.1 with a period of 190 yr. At the same time, $a_{\rm{c}}$ and $e_{\rm{c}}$ may range from 1.368 to 1.393 AU and from 0.03 to 0.06, respectively.  The period ratio based on the Newtonian model is $P_{\rm{c}}/P_{\rm{b}}=0.349$, so it is even closer to the 7:20 commensurability. The eccentricity-type resonant angles, defined as a linear combination of mean longitudes and arguments of periastron, show no libration. The difference between arguments of periastron, defined as $\Delta \omega = \omega_{\rm{c}}-\omega_{\rm{b}}$ demonstrates a lack of apsidal alignment and oscillations. Thus, the system is not in a dynamical resonance. Variations in eccentricities in the first thousand years are shown in Fig.~\ref{Fig.DynVariations}.

\begin{table}
\centering
\caption{Dynamical orbital parameters of TYC 1422-614-1 b and c.  \label{Tab.DynResults}}
\begin{tabular}{lll}
\hline
Parameter & TYC 1422-614-1 b & TYC 1422-614-1 c \\
\hline
\hline
$P$ (days)                                        & $198.44 \pm 0.64$ & $569.2 \pm 2.1$ \\
$T_0$ (MJD)                                       & $52842 \pm 40$      & $52616 \pm 30$ \\
$K$ ($\textrm{m s}^{-1}$)                         & $82.2 \pm 3.7$     & $232.8 \pm 3.3$ \\
$e$                                               & $0.07 \pm 0.04$   & $0.049 \pm 0.014$ \\
$\omega$ (deg)                                    & $62 \pm 46$       & $124 \pm 20$ \\
$m_2\sin i$ ($\textrm{M}_{\textrm{J}}$)           & $2.51 \pm 0.12$           & $10.10  \pm 0.14$ \\
$a$ (AU)                                          & $0.6879 \pm 0.0015$          & $1.3916 \pm 0.0033$ \\
$V_0$ ($\textrm{m s}^{-1}$)                       & \multicolumn{2}{c}{ $-67.6\pm3.7$ } \\
$\sqrt{\chi_\nu^2}$                               & \multicolumn{2}{c}{ $1.59$ } \\
$\sigma_{\textrm{RV}}$ ($\textrm{m s}^{-1}$)      & \multicolumn{2}{c}{ $17.17$ } \\
$N_{\textrm{obs}}$                                & \multicolumn{2}{c}{ $86$ }\\
\hline
\end{tabular}
\label{NewtonianFit}
\end{table}

\section{Stellar activity}\label{activity}

Giants are well known to exhibit all sorts of variations, stellar
activity, pulsations and rotation-induced effects should be considered
before putting forward a companion hypothesis.

\subsection{Ca II H$\&$K lines}

The flux at the cores of the Ca II H \& K lines is a good tracer of
chromospheric activity (e.g., \citealt{ Noyes1984, Duncan1991}) and, 
therefore, the Ca II H \&K line profiles are widely accepted as stellar activity indicators.
Both the Ca~II H \& K lines and the infrared Ca~II triplet lines at
849.8-854.2 nm lay outside HET \& HRS wavelength range but the Ca~II H
\& K lines are available to us in the TNG HARPS-N spectra. 
The SNR in the blue spectrum of our red giant star is not very high, 
($\sim$10-20), but no trace
of reversal, typical for active stars
\citep{EberhardSchwarzschild1913} is present, and thus no obvious
chromospheric activity can be demonstrated.

 To quantify possible activity induced line profile variations we
 calculated an instrumental S$_{HK}^{inst}$ index according to the 
 prescription of \cite{Duncan1991} for all 17 epochs of HARPS-N
 observations and correlated them with RV and BS. 
 The mean value of $\overline {S_{HK}^{inst}}$=0.25 $\pm$ 0.03 which
 places TYC 1422-614-1 rather among non-active subgiants according to
 \cite{IsaacsonFischer2010}. 
 The correlation coefficient r=0.56 is below the critical value
 (r$_{15,0.01}$=0.61) and no correlation between S$_{HK}^{inst}$ and
 RV can be inferred from our data. There is also no relation between
 S$_{HK}^{inst}$ and BS (r=0.52).  Much better relation exists between
 our S$_{HK}^{inst}$ index and the RV  uncertainty (r=-0.65) which
 suggests that the observed S$_{HK}^{inst}$ scatter in our data is
 actually only noise and that no measurable chromospheric activity is
 present in TYC 1422-614-1.

\subsection{$H\alpha$ analysis}

 \begin{figure}[p]
   \centering
 \includegraphics[width=0.75\textwidth]{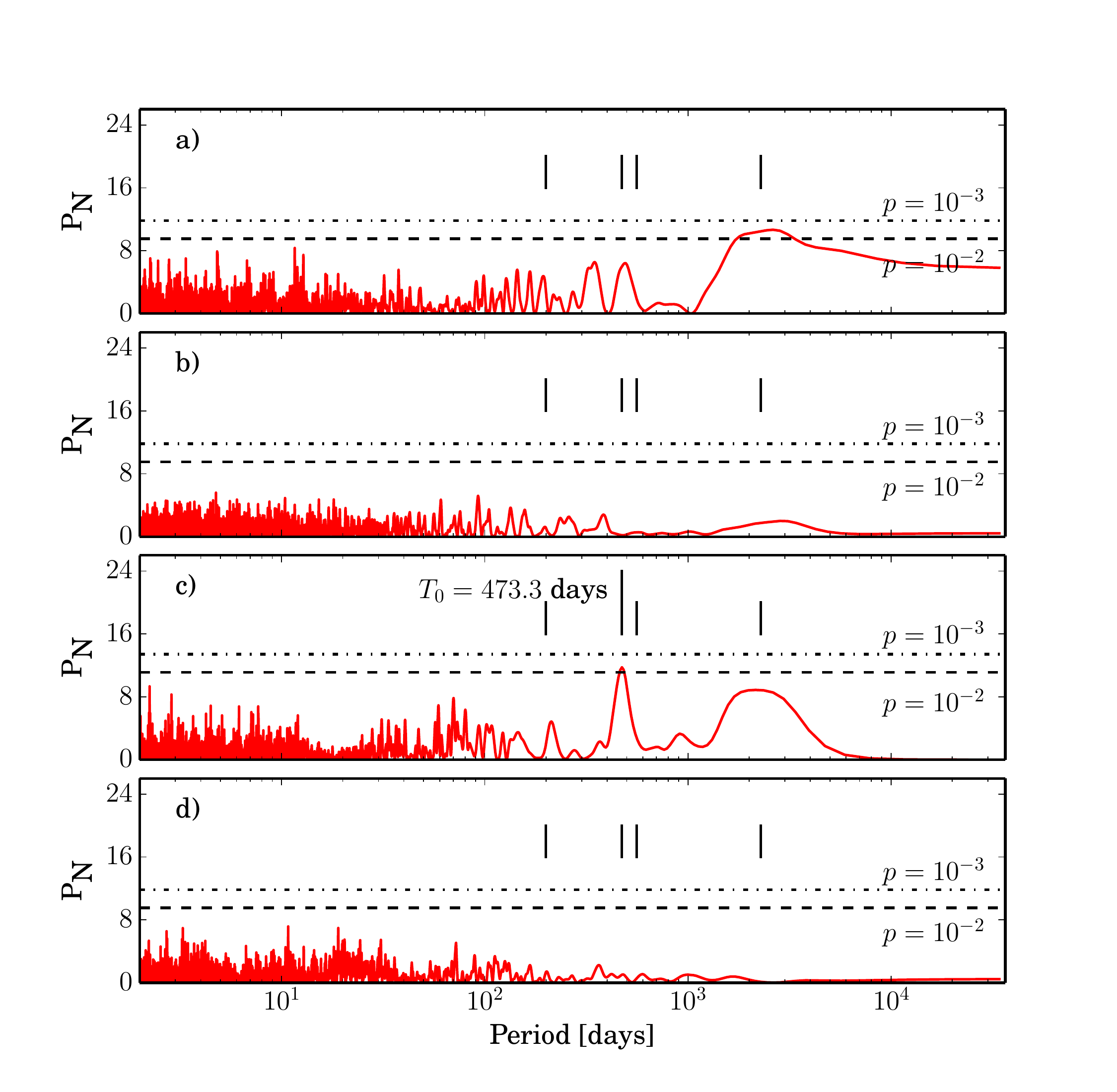}
      \caption{From top to bottom Lomb-Scargle periodograms of (a) $I_{H\alpha}$ activity
        index, (b) $I_{Fe}$ index, (c) ASAS photometry, and (d)
      HET \& HRS BS. See also Figure \ref{LS}.}
         \label{LS-ph}
         \end{figure}

In addition to  the Ca~II H \& K lines we use the HET \& HRS H$\alpha$ (656.2808 nm) line data as a chromospheric activity indicator
since, as \cite{2007A&A...469..309C} showed, the calcium and hydrogen
lines indices do not always correlate and cannot be used
interchangeably as activity indicators. 

Following the procedure described in detail in
\cite{2013AJ....146..147M} which is based on the approach presented by
\cite{2012A&A...541A...9G} and \citet[][and references
therein]{2013ApJ...764....3R}, we measured the H$\alpha$ index
($I_{H\alpha}$) for the 69 HET \& HRS spectra. To take possible instrumental effects into
account we also measured the index
of the insensitive to stellar activity Fe~I 654.6239 nm control line
($I_{Fe}$). Moreover, as the wavelength regime relevant to H$\alpha$ and
Fe~I line indices may still contain weak $\mathrm{I_{2}}$ lines, we
also measured H$\alpha$ and Fe~I indices for the iodine flat-field
spectra ($I_{\mathrm{I_{2}}, H\alpha}$ and $I_{\mathrm{I_{2}}, {Fe}}$
respectively). 

The marginal rms variations of the $I_{\mathrm{I_{2}}, H\alpha}$ =
0.11\% and $I_{\mathrm{I_{2}}, {Fe}}$ = 0.32\% in comparison to the
relative scatter of the $I_{H\alpha}$ = 2.19\% and the $I_{Fe}$ =
1.1\% index assure us of the negligible contribution of the weak
iodine lines to the H$\alpha$ and Fe line indices in TYC 1422-614-1. 
The significant relative scatter variation of the Fe line equal to
half of the H$\alpha$ rms variation is driven by two outliers
exceeding $\pm 3\sigma$ level from the mean value and drop to 0.64\%
after excluding them.  

There are no significant signals present in the LS periodograms of the
$I_{Fe}$ index (Figure~\ref{LS-ph}b), in contrast to the $I_{H\alpha}$
periodogram, where a significant signal with a period of
$\sim$2550 days (almost 7 years), similar to that observed in the RV
residuals, is clearly visible (Figure~\ref{LS-ph}a). Therefore we put
forward the hypothesis that this periodicity is driven by long-term
chromospheric activity of TYC 1422-614-1 analogue to the eleven-year
solar cycle. 

No excess fluctuation power in the LS periodogram of the
$I_{H\alpha}$ index is present at the 198.44 and 569.2 day periods
detected in the RV data. We therefore conclude that none of the signal present in radial
velocity data is driven by chromospheric activity of the parent star.

\subsection{Photometric variability \label{SecPhotom}}

From ASAS \citep{Pojmanski2002} we have available 367 epochs of TYC
1422-614-1 observations covering  HJD 2452622 and 2454989 and thus, partly contemporaneous with
out HET observations.   
The photometric observations were compiled from 4 different fields,
two of which contain enough data for independent analysis with 135
(Field 1) and 185 (Field 2) epochs, respectively. 
After iterative 3$\sigma$ filtering 355  epochs can be used for
photometric variability analysis. These data show mean brightness of
m$_{ASAS}$=10.186$\pm$0.011 mag (rms).   The scatter in brightness of
TYC 1422-614-1 is much larger than the p-mode oscillations amplitude
predicted from the scaling relations of \cite{KjeldsenBedding1995}, 
0.13 mmag (the reason obviously being the limited precision of ASAS
photometry for this star). The star was not classified as a variable
in ASAS-3. 

The Field 1 data show no significant periodic signal, but both Field 2
and all combined data show a weak periodic signal of 473 days (see Figure \ref{LS-ph}c).
 To check whether this periodic signal is real we performed a
 bootstrap analysis and after 10\,000 trials we found
 FAP$_{10,000}$=0.02 so although weak, it cannot be ignored. 
A formal sine fit to combined data with a period of 473 days results
in an amplitude of 0.011 $\pm$ 0.002 mag, i.e. at the level of  the mean
brightness uncertainty. 

What might be the origin of that periodic signal? It is much longer than those of
p-mode oscillations ($\sim$3 h) and radial pulsations (0.52d). It is
also well outside the uncertainties of the two Keplerian periods found in
Sections \ref{SecKepler} and \ref{SecNewton}. 
 
The only source of such photometric periodicity  (if indeed real) may
be a spot or a group of spots rotating with the star and therefore it
might represent the true stellar rotation period. 
Using the stellar radius from Table \ref{Parameters} we obtain a
v$_{rot}$=0.7 km s$^{-1}$ suggesting a very slow rotation rate.  

The value of the projected  rotation velocity obtained  by \cite{Nowak2012}
 (see Table \ref{Parameters}) is lower than the average K giant
projected rotational velocity of 3 km s$^{-1}$ \citep{Fekel1997}. This suggests that we are dealing with a very slow rotator
and as such its measured rotational speed value is expected to be
uncertain as the projected rotation velocity 
is $\sim$3 times lower than the instrumental PSF. 
If we assume that the 473 days photometric period is the stellar
rotational period then we find good agrement, within uncertainties,  between the photometric
rotation period  and the value obtained by \cite{Nowak2012} from
spectral analysis, $v_{rot}^{CCF}$ $\sin
i_{\star}$= 1.4$\pm$0.7  km s$^{-1}$. 
A slow rotational velocity for TYC 1422-614-1 is as well consistent 
with the star having low chromospheric activity (see e. g.
\citealt{doNascimento2003}). 
We therefore assume that the rotation period of TYC 1422-614-1 is 473 days.

We can use the ASAS data on TYC 1422-614-1 to estimate an upper limit to the stellar
surface covered by hypothetic spots.    
Under the assumption that the  only  source  of observed brightness scatter are
spots rotating on the surface of TYC 1422-614-1 the fraction, $f$, of the
star covered is then
determined from the mean brightness change (mean brightness
uncertainty) and we obtain a $f\approx$1.5$\%$. With it  and the rotational 
period  derived from ASAS photometry, we consider the
impact of such spots on the RV and BS as they can alter spectral lines
profiles, and mimic doppler shifts \citep{Walker1992}.  
Using the results of  the extensive simulations of \cite{Hatzes2002} we
estimate that  spots of f$\approx$1.5$\%$  will result in RV and
BS variations with an amplitude of 20 m s$^{-1}$ and 27 m s$^{-1}$,
respectively at maximum.   

The observed RV variations in TYC 1422-614-1 are 15 times larger, so they cannot
originate from spectral line deformations due to spots. Furthermore,  
any contribution to the observed RV variations from stellar spots is expected to
be very small, at the level of $\sim$7$\%$. Our assumption that the
observed RV variations are due to Doppler shifts is therefore fully
justified.  We note, however, that the ASAS photometric data show a trace
of a long period similar to that present in RV residua and
$I_{H\alpha}$. 

The observed BS variations show a similar (HARPS-N) or
even larger (HET) amplitude than the estimated one, it is therefore of special importance to
study in more detail the BS.

\subsection{Line bisector analysis}

The HET\& HRS and HARPS-N BS are calculated from different instruments
and sets of spectral lines. They are not directly comparable and so we
consider them separately.

The HET\& HRS BS LS periodogram shows no significant periodic signal
(Figure \ref{LS}). Although the observed BS variations are 
larger than the values estimated from photometry there is no correlation between
RV and BS. Altogether, in the HET data, there is no observational evidence of 
spectral line deformation influence on RV. The observed HET BS
scatter may reflect either additional stellar activity that we do not
resolve in time with our observations, like p-mode oscillations,  or be 
the effect of underestimated uncertainties. 

The TNG HARPS-N data is more precise than the HET \& HRS but we do
not have enough epochs of observations for a conclusive periodogram analysis.  
The HARPS-N BS amplitude is comparable to the HET \& HRS BS standard deviation.
The RV and BS show a correlation of r=0.63 which is marginally
significant as the critical value is r$_{15,0.01}$=0.61. A closer look
at the BS data shows however, that the correlation may very well be a
random effect due to small statistics, as variations of $\sim$16 m
s$^{-1}$ were observed over one night.  
Nevertheless,  the amplitude of observed BS variation is consistent
with the upper limit estimated from the available photometry. 

Fortunately enough the HARPS-N data cover the full observed RV amplitude,
and the amplitude of BS variations  supports our conclusion from
Section \ref{SecPhotom} that  the spectral line deformation due to  a
hypothetical spot, if real,  may contribute only to a minor fraction
of the observed RV variations. 

We conclude, therefore, that although the HET \& HRS BS data show no
evidence of a stellar activity influence on the measured RV,  the HARPS-N BS
data prove a lack of chromospheric activity but suggests an additional RV
signal from a spectral line distortion of yet unknown  period. The  RV
scatter present in post fit residua is too large for unresolved p-mode
oscillations, but it is consistent with our estimates of the RV
contamination due to spots.

\section{Discussion\label{discussion}}

\begin{figure}
   \centering
   \includegraphics[width=0.75\textwidth]{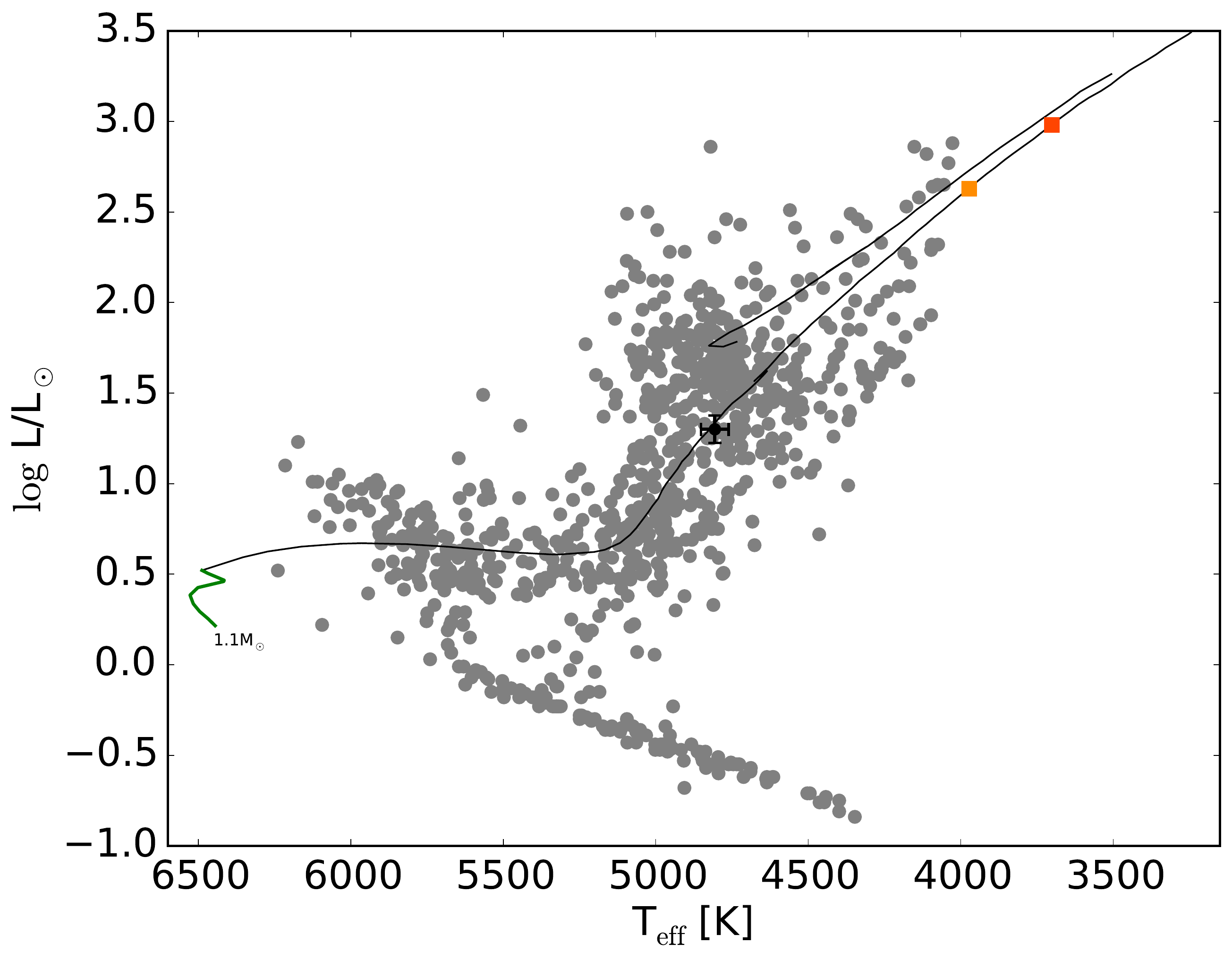}
      \caption{Hertzsprung-Russel diagram for the complete PTPS sample with TYC 1422-614-1 position indicated and 1.1 M$_{\odot}$, z=0.008  star  evolutionary track from Bertelli et al. (2008) and,  in green, the path of the star with planet c being within the optimistic HZ of \cite{Kopparapu2013}.  The orange and red rectangles indicate the phase of  planet b  and c ingestion, respectively.}
         \label{HRD}
         \end{figure} 

Both the stellar spectroscopic parameters and the stellar luminosity converge to locate TYC 1422-614-1 in the track of a  K2 giant quickly evolving up the Red Giant Branch (RGB). The star is already
undergoing the
first dredge-up but, within uncertainties, it is most likely located
before the luminosity jump  (see Figure
\ref{HRD}). The radius of the star is estimated to be 6.85 $\pm1.38$
R$_{\odot}$. 

Considering angular momentum conservation alone,  the
innermost planet located at 0.69 \,AU (and with a/R$_*$ = 22) is
not expected to have experienced orbital decay
caused by the stellar tides nor orbital expansion due to mass-loss
(see \citealt{VillaverLivio2009} and \citealt{Villaveretal14}).  The planet, 
however, is expected to be engulfed by the star as it ascends to the tip of the RGB. Using the approximate location of TYC 1422-614-1 following a
path of a M=1.1 M$_{\odot}$ z=0.008 star from \cite{Bertelli2008} we
calculated the evolution of the 
planet orbit and we find that it is expected to  be engulfed in all
the possible scenarios considered under several
mass-loss prescriptions. Our orbital evolution calculations
show that planet $b$ is expected to enter the stellar envelope as the
planet orbit decays due to tidal forces when the star reaches a radius
of $\approx$43 R$_{\odot}$. 
An estimated upper limit for the time left
for the planet before entering the stellar envelope is $\approx$120 mln yr (computed assuming the
star follows the \citealt{Bertelli2008} track), 
during its ascend on the RGB. 

The outermost planet is
expected to reach the stellar surface as well, 
but a bit later, when the
stellar radius reaches a value of $\approx$75 R$_{\odot}$ in about 130 mln yr. While the innermost planet
will be most likely destroyed inside the stellar envelope (see
\citealt{VillaverLivio2007}), the outcome for the outermost planet is more uncertain as, given its large
mass, might provide enough angular momentum at the stellar surface to
trigger the partial ejection of the envelope
\citep{Garciasegura2014} and/or get partially eroded and become a close companion to the star 
\citep{Bear2011, Bear2012}.

\begin{figure}
   \centering
   \includegraphics[width=0.75\textwidth]{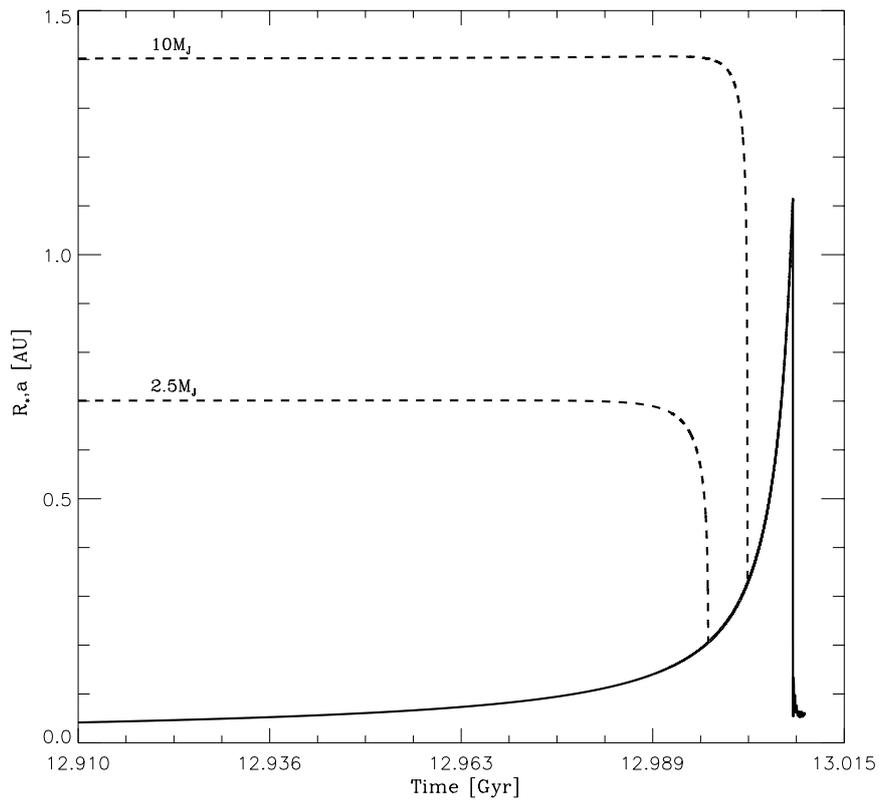}
      \caption{The solid line represents the last 100 Myr in the evolution of the
  stellar radius as the star evolves to the tip of the RGB. The star is evolving under the Reimers mass-loss
  prescription with $\eta=0.6$. The evolution of the initial orbits of
  planets TYC 14221614-1 b and c are shown as dashed lines asuming their
  M$_p sin i$  minimum masses. }
         \label{jupiters}
         \end{figure}

The evolution of the stellar radius over the last 100 Myr along the RGB
phase is shown in Figure \ref{jupiters} together with the expected orbital
evolution of the planets in the TYC 1422-614-1  system. Note, that we have
assumed the minimum planet mass for the calculation and we show a star
that undergoes Reimers mass-loss prescription with an $\eta=0.6$
 (for further details see \citealt{VillaverLivio2009,Villaveretal14}).

If we consider the evolutionary status of the star in terms of the
stellar surface gravity, TYC 1422-614-1 with  logg$=2.85$ is the second
most evolved multiple planetary system found yet around a giant star
(the record holder is BD+20 2457 with its logg$=1.77$). In this
regard, BD+20 2457 and TYC 1422-614-1 are very similar and interesting systems: both host a multiple
planetary system around a very evolved star (R/R$_{\odot}$= 33 and
6.85 respectively) with mass close to solar and including a massive
planet/BD  M$_p >$ 10 M$_J$.

Assuming its current
orbital separation, planet $c$ was within  the ''optimistic'' Habitable
Zone (HZ) defined by \cite{Kopparapu2013} for about 4 billion  years
after the star reached ZAMS. We have then a solar mass star with a long
MS evolution and a giant planet/BD that stayed in the HZ for a
reasonable time. 
 
Currently, due to the evolution of the location
of the HZ as the star leaves the MS neither planet $b$ nor $c$  reside
within HZ (see i. e. \citealt{Lopez2005, 
  Danchi2013}). 

TYC 1422-614-1 belongs to the very small sample\footnote{exoplanet.eu, exoplanet.org} 
of known multiple planetary systems around giants (i.e., stars with logg $<$ 3.5). Multiple
systems around giants include:  
BD+20 2457 b, c  (\citealt{Niedzielski2009b}); 
24 Sex b,c and HD 200694 b, c (\citealt{Johnson2011}); 
HD 4732 b, c (\citealt{Sato2013});
Kepler 391 b, c \citep{Rowe2014}; 
and Kepler 56 b, c, d 
\citep{2013Sci...342..331H}
and represent
$\approx$16 $\%$ of all the planets orbiting giant stars (including TYC 1422-614-1). This fraction is 
slightly smaller, but consistent, with that on the MS considering that
about $\approx 20\%$ planets
orbiting MS stars are in multiple systems. 

The two companions orbiting TYC 1422-614-1 have semimajor axes of 0.69
and 1.37 AU and minimum masses of 2.5 and 10.0 M$_J$. The outer
companion has a mass  that is suspiciously close to the upper
planet mass limit for planets and as such its plausible formation
mechanism justify further discussion. 

There are only  five  other 
multiple systems known containing a  likely BD-mass companion: 
HD 168443 b, c - \cite{Marcy2001}; 
HD 38529 b, c - \cite{Fischer2003}; 
HD 202206 b,c - \cite{Correia2005}; 
HAT-P-13 b, c - \cite{Bakos2009}; 
BD+20 2457 b, c  (\citealt{Niedzielski2009b}); 
and all of them with the exception of BD+20 2457 
and HD 202206  have the more massive component in the outermost
orbit. Apparently such systems exist very scarcely around solar-mass stars of
various metallicity and evolutionary stage  but it is important to
note that 2 out of the 6 known
revolve around giants.  Furthermore, the system of TYC 1422-614-1 presents almost
circular orbits while in other systems of this group huge
eccentricities of up to 0.662 (HAT-P-13 c) are common. 

The detection of an object with a minimum mass close to the BD limit in a relatively close orbit to a solar mass
star, may offer additional clues to the yet unknown most likely
formation mechanism of BD and its relation 
with star and planet formation. In fact, the existence of the BD desert \citep{Marcy2000,Grether1996},  the observed 
paucity of BDs companions to solar-mass stars
within 3 AU, has been put forward
as possible evidence of a differentiation in the formation mechanism
according to the object mass.  

In TYC 1422-614-1, like in 4 out of the
6 in total systems of the class, the smaller mass planet is in an
innermost orbit. Because the outer planet is clearly
the more massive, it is unlikely that the inner planet
scattered the outer planet to its current orbit.  
Furthermore, if we assume that the outer planet in these systems
formed in the stellar
protoplanetary disk it is reasonable  to assume that the inner planet
also formed there. In fact, theoretically ``planets'' with masses $> 10 M_J$ can
in principle be formed by core-acretion in protoplanetary disks
\citep{Mordasini2009}. However, it is
interesting to note that the metallicity distribution of BDs
companions seems inconsistent with the predictions of the
core-acretion formation mechanism \citep{Ma2014} and that TYC 1422-614-1
has indeed sub-solar metallicity. 

On the other hand, we can consider that the largest planet/BD in these
systems is formed via the alternative mecanism: disk
fragmentation (see e.g. \citealt{Stamatellos2009}). Most of the BD
formed by this mechanism are expected to be either ejected from the
system, or stay bound to the central star at relatively wide
orbits. Thus, a closely bound system such as TYC 1422-614-1 is a very unlikely outcome
of this process. Furthermore, the formation and location of the
innermost planet raises even further questions on the possibility of disk
fragmentation as the formation mechanism for the massive companion.  

Two other multiple systems with a massive planet/BD orbiting close
to the star: HD 202206 \citep{Correia2005} and BD+20 2457
\citep{Niedzielski2009b} have architectures in which the innermost
companion is the more massive one, suggesting that if it formed as a BD 
by disk instability the second outermost planet might have formed in
the circumbinary disk composed by the main star and the BD
(see e.g. \citealt{Correia2005}).

All together the rare class of objects to which TYC 1422-614-1 belongs
might provide important clues into the planet/BD formation
mechanism and limits
by setting restrictions into the smallest mass clump that the disk
fragmentation mechanism allows or the largest mass that core acretion
is capable.  While disk fragmentation seems to be
the most likely mechanism for BD formation in general, the
presence of an innermost planet and the close orbit of the massive
companion raises some important questions on the viability of this
mechanism for the formation of systems like TYC 1422-614-1. The core
accretion model, given the low metallicity of the star could be
problematic as well unless a much more massive disk than usual is assumed to increase the
efficiency of the process.

\section{Conclusions \label{conclusions}}

Our combined set of observations of TYC 1422-614-1 obtained with HET \& HRS and TNG \& 
HARPS-N show high precision RV variations that can be attributed with
a high level of confidence to  Doppler shifts. Our analysis of the
data combined with photometric time-series from  ASAS lead us to propose
the discovery of two low-mass components orbiting that K2 giant.  
We find an additional weak, long-term component in the post-fit RV residue, 
$I_{H\alpha}$ and ASAS photometry which we argue is most likely due to stellar activity or
yet another companion. A stellar rotation period of 473 days is
determined from ASAS photometry which is in agreement with results of
spectroscopic determination.  

The planetary system orbiting TYC 1422-614-1 is one of a very special
kind, as the star is a solar mass giant with multiple planets (only
other 5 such systems are known), the star is in an
advance evolutionary stage up the RGB, and it is hosting a planet with a
minimim mass close to the BD limit in a 1.37 AU orbit. The existence of
such a massive planet so close to the star can offer important
constraints to the planet/BD formation mechanims and limits as the
star has sub-solar abundances and another relative massive planet is
orbiting the star in an innnermost location. Furthermore, although the
HZ is expected to have moved outwards as the star left the MS, it is
also interesting to note that TYC 1422-614-1 b could have remained in
the HZ for about 4 billion years. Finally, following orbital evolution
both planets are expected to reach the stellar surface long before the
star finalized its ascend up the RGB due to orbital decay caused by
tidal interaction.

\section{Acknowledgements}

We thank the HET  and TNG  resident astronomers and telescope operators  for support.

AN, MoA, KK, GN, BD and MiA  were supported in part by the Polish Ministry of Science and Higher Education grant N N203 510938 and by the Polish National Science Centre grant UMO-2012/07/B/ST9/04415.
MoA was also supported by the Polish National Science Centre grant no. UMO-2012/05/N/ST9/03836.
EV work was supported by the Spanish Ministerio de Ciencia e Innovacion (MICINN), Plan Nacional de Astronomia y Astrofisica, under grant AYA2010-20630 and by the Marie Curie grant FP7-People-RG268111. 
AW was supported by the NASA grant NNX09AB36G. 
GM acknowledges the financial support from the Polish Ministry of Science and Higher Education through the Iuventus Plus grant IP2011 031971.
D.A.G.H. acknowledges support for this work provided by the Spanish Ministry of Economy and Competitiveness under grant AYA-2011-29060.

The HET is a joint project of the University of Texas at Austin, the Pennsylvania State University, Stanford University, Ludwig-
Maximilians-Universit\"at M\"unchen, and Georg-August-Universit\"at G\"ottingen.
The HET is named in honor of its principal benefactors, William P. Hobby and Robert E. Eberly.

The Center for Exoplanets and Habitable Worlds is supported by the Pennsylvania State University, the Eberly College of Science, and the Pennsylvania Space Grant Consortium. 

This research has made use of the SIMBAD database, operated at CDS (Strasbourg, France) and NASA's Astrophysics Data System Bibliographic Services.

This research has made use of the Exoplanet Orbit Database and the Exoplanet Data Explorer at exoplanets.org.




\end{document}